\documentclass[a4paper,fleqn]{cas-dc}

\usepackage[numbers]{natbib}

\usepackage{amsmath}
\usepackage{amssymb}
\usepackage{booktabs}
\usepackage{array}
\usepackage{tikz}
\usepackage{pgfplots}

\usepgfplotslibrary{groupplots}

\pgfplotsset{compat=1.18}

\usepackage{multirow}
\usepackage{graphicx}
\usepackage{svg}
\usepackage{algorithm}
\usepackage{algorithm}
\usepackage{algpseudocode}

\begin{document}

\let\WriteBookmarks\relax
\def\floatpagepagefraction{1}
\def\textpagefraction{.001}

% ------------------------------------------------
% SHORT TITLE / SHORT AUTHORS
% ------------------------------------------------

\shorttitle{SynCo: Synthetic Community-Aware Attributed Graph Generator}

\shortauthors{Messias et al.}

% ------------------------------------------------
% TITLE
% ------------------------------------------------

\title[mode=title]
{SynCo: Synthetic Community-Aware Attributed Graph Generator
for Graph Neural Network Benchmarking}

% ------------------------------------------------
% AUTHORS
% ------------------------------------------------

\author[1]{Guilherme Henrique Messias}[orcid=0009-0000-6820-2205]
\ead{ghmessias@estudante.ufscar.br}

\cormark[1]

\author[2]{Mariana Caravanti de Souza}[orcid=0000-0002-1746-8414]
\ead{mariana.caravanti@ufms.br}

\author[1]{Sylvia Iasulaitis}[orcid=0000-0002-3526-1003]
\ead{si@ufscar.br}

\author[1]{Alan Demétrius Baria Valejo}[orcid=0000-0002-9046-9499]
\ead{alanvalejo@ufscar.br}

\cortext[cor1]{Corresponding author}

% ------------------------------------------------
% AFFILIATIONS
% ------------------------------------------------

\affiliation[1]{
    organization={Federal University of São Carlos},
    city={São Carlos},
    state={São Paulo},
    country={Brazil}
}

\affiliation[2]{organization={Federal University of Mato Grosso do Sul},
city={Campo Grande},
state={Mato Grosso do Sul},
country={Brazil}
}

% ------------------------------------------------
% ABSTRACT
% ------------------------------------------------

\begin{abstract}
Graph Neural Networks (GNNs) are powerful models for handling attributed graphs in tasks such as classification, link prediction, and community detection, as they enable the aggregation of information from both structural and semantic sources. However, progress in community detection is hindered by the lack of high-quality datasets, since ground-truth community labels are often unavailable and most algorithms proposed in recent literature rely on the same benchmark datasets for model training and evaluation. To address this issue, attributed random graph generators are commonly employed to create synthetic graphs for assessing the strengths and limitations of GNN-based models. Nevertheless, most existing generators rely heavily on power-law degree distributions, despite recent evidence indicating that scale-free networks are rare, particularly in social network contexts. Moreover, state-of-the-art attributed graph generators provide limited flexibility, as they do not allow users to construct communities with varying densities, degree distributions, and sub-community structures. To overcome these limitations, we introduce the Synthetic Community-Aware Attributed Graph Generator (SynCo), a graph generation algorithm that allows users to control the node degree distribution and sub-community structure. We evaluate SynCo across three different tasks: graph mimicking, hyperparameter evaluation, and node clustering tuning. The results show that our model outperforms state-of-the-art approaches in synthetic graph generation and data augmentation, while preserving the original distributions of duplicated and augmented datasets, as confirmed by statistical tests well know in literature. We also demonstrate the ability of SynCo to generate nodes in large scale, up to 2.1 million nodes.
\end{abstract}

% ------------------------------------------------
% HIGHLIGHTS
% ------------------------------------------------

% \begin{highlights}

% \item SynCo generates attributed graphs with explicit community and
% sub-community structure.

% \item The framework supports controlled graph generation, graph
% mimicking, and graph augmentation.

% \item SynCo enables explicit control over degree distributions,
% heterophily, structural noise, and semantic noise.

% \item Experiments demonstrate its use for stress-testing node
% clustering algorithms under controlled scenarios.

% \end{highlights}

% ------------------------------------------------
% KEYWORDS
% ------------------------------------------------

\begin{keywords}

Attributed graph generation
\sep Graph neural networks
\sep Community detection
\sep Node clustering
\sep Synthetic graphs
\sep Graph augmentation

\end{keywords}

\maketitle

% ------------------------------------------------
% PAPER
% ------------------------------------------------

\section{Introduction}
\label{sec:introduction}

Community detection in attributed graphs is a fundamental task in network science and machine learning, with applications ranging from social network analysis \cite{chandrika2022graph} to biological systems \cite{zhuo2022model}. The evaluation of algorithms in this domain, however, is highly dependent on the availability of suitable datasets, and the performance of state-of-the-art models has not yet reached a stage where the results are equal to ground truth partitions \cite{wei2024overview}. Furthermore, community detection requires this ground-truth information to reflect and evaluate group structures, yet such labeled datasets are scarce and often difficult to obtain \cite{maekawa2023gencat}. This scarcity has motivated the research community to rely on datasets originally designed for classification tasks as a proxy for evaluating clustering and community detection methods \cite{Messias2025systematic, liu2026bridging}.

Although convenient, the use of classification datasets for this purpose is problematic. Many of these datasets do not exhibit inherent community-like structures, which undermines their suitability for assessing algorithms aimed at uncovering cohesive groups in networks \cite{lai2023reevaluation}. These limitations highlight the pressing need for synthetic attributed graph generators that better capture the properties required to evaluate community detection approaches \cite{zhou2025data}.

To overcome these challenges, several algorithms have been developed to generate synthetic graphs tailored for evaluation. Some approaches rely on latent factors to construct networks \cite{maekawa2019acmark}, while others employ traditional models such as the Barabási–Albert \cite{Andreeva2023}  model or preferential attachment \cite{Benyahia2016dancer} to reproduce well-known structural properties. In addition, certain methods enable the mimicry of existing datasets, thereby functioning as data augmentation strategies for training Graph Neural Networks (GNNs) \cite{Wassington2024skymap, maekawa2023gencat}. These efforts have contributed to alleviating the scarcity of benchmark data, but they still face significant limitations, particularly in their inability to flexibly generate datasets where each community can exhibit distinct densities, node counts, and edge distributions. Moreover, when mimicry is applied, most approaches struggle to preserve the original node degree distribution, which undermines their capacity to reproduce realistic structural patterns necessary for robust evaluation of community detection algorithms.

To address these limitations, we propose a Synthetic Community-Aware Attributed Graph Generator (SynCo), a model capable of generating attributed graphs with different cluster densities, node counts, edge distributions, and controlled homogeneity. The proposed framework is composed of four main stages: Node community and sub-community assignment, Edge construction, Noise generation, and Attribute generation. With this, the proposed model is capable of (1) generating artificial datasets according to user-specified distributions via matrices that define intra- and inter-community relationships, (2) mimicking existing datasets while preserving their structural and attribute characteristics, and (3) augmenting a given mimicked graph. To assess these capabilities, we evaluate SynCo in three tasks: Parametric evaluation in synthetic data generation, scale-free analysis over mimicked data according to \cite{Broido2019scale} and \cite{Clauset2009Power}, and a stress scenario over 10 different node clustering algorithms, enabling an understanding of the limitations of these algorithms under varying numbers of clusters, heterogeneity, and structural or semantic noise.

Our main contributions can be summarized as follows:
\begin{itemize}
    \item We introduce a novel model for artificial attributed graph generation based on sub-community structures. This model allows the creation of graphs with flexible community-level characteristics, such as varying densities and distinct intra-cluster edge distributions.

    \item A model that is also capable of dataset mimicking, maintaining local structures of sub-communities and being able to augment the original graph, also serving as a graph augmentation algorithm.

    \item We conduct a comprehensive evaluation of state-of-the-art graph generation models, considering both the generation of synthetic datasets for benchmarking community detection with GNN-based approaches and the feasibility of dataset mimicking while maintaining degree distributions. This dual evaluation highlights the advantages of SynCo in producing more realistic and diverse attributed graphs for assessing community detection algorithms.
    
\end{itemize}

The remainder of this paper is organized as follows. Section~2 presents the problem statement and reviews related work on synthetic attributed graph generation. Section~3 describes the proposed SynCo model and its generation mechanisms. Section~4 details the experimental setup and empirical evaluation. Finally, Section~5 discusses the main findings and concludes the paper.

\section{Problem Statement and Related Works}

In this section we present the problem statement and the related works in synthetic graph generation. 

\subsection{Problem Statement}
\label{sec:problem_statement}

Synthetic attributed graph generation addresses the problem of constructing graphs with both structural and attribute information under different modeling assumptions. In general, existing approaches can be grouped into two broad problem settings. In the first setting, the goal is to generate a graph from a set of user-specified constraints, such as the number of nodes and edges, degree distributions, or attribute assignment mechanisms. In the second setting, the objective is to generate a synthetic graph that resembles a given real graph, preserving selected structural and attribute-related properties.

Formally, in the first setting, given a set of user inputs $\mathcal{S}$, including, for instance, the number of nodes, the number of edges, degree distributions, and probability assignment matrices, the goal is to generate a graph $G = (V, E, X, Y)$ that satisfies the specified constraints. In the second setting, given an input graph $G = (V, E, X, Y)$, the objective is to construct a synthetic graph $\tilde{G} = (\tilde{V}, \tilde{E}, \tilde{X}, \tilde{Y})$ that preserves relevant structural and attribute-related properties of the original graph.

\subsection{State-of-the-art in Attributed Graph Generators}

DANCer \cite{Benyahia2016dancer} is a dynamic attributed network with community structure generator, that leans on properties of real-world networks such preferential attachment, small world and homophily. The network generation consists of two phases: (1) Built an initial graph with the well-know network properties, and (2) modify the initial graph with micro (removing or adding nodes and edges) and macro operations (migrating members of a community to a new community or existing one, splitting communities and merging then).

GenCAT \cite{maekawa2023gencat} generates graphs from user-specified inputs that control the relationship between node attributes and labels, enabling the modeling of both core and border nodes, as well as homophilic and heterophilic interaction patterns. The framework is scalable and flexible in terms of structural and semantic configurations. 
% However, despite its usefulness, GenCAT restricts the node degree distribution to power-law regimes, limiting its ability to generate graphs following alternative degree distributions.

The node degree distribution (NDD) encodes both local and global structural properties of a graph, providing essential insights into its connectivity patterns and overall organization. In this context, \cite{miller2011efficient} proposed a computationally efficient random graph model capable of generating synthetic networks from a known NDD. By preserving the expected degree sequence, the model captures fundamental structural characteristics of real-world graphs, making it particularly suitable for graph mimicking and for producing synthetic networks with high structural fidelity to empirical data.

To address the limitations of GenCAT, acMARK \cite{maekawa2019acmark} allows the generation of graphs in normal and uniform distributions. The algorithm also make use of latent factors to generate topological relations and feature relations over the examples, however, it allow only homophily graphs, which are more suitable for community detection tasks. The graph generation is evaluated using benchmarks for community detection over different topologies created using different values of parameters.

Although homophily graphs are more suitable for clustering, real-world networks typically have heterophilic characteristics. AL-BTER allow the generation of graphs with low assortativity, which represents more challenge on graph learning tasks. Besides low node assortativity, the model allows the generation of low attribute assortativity, i.e, similar attributed nodes (measured by cosine similarity) tend to have different classes. To guarantee the quality of the proposed approach, AL-BTER was evaluated on synthetic graph properties and tasks such as classification using GNNs.

SkyMap \cite{Wassington2024skymap} is a generative model designed to create synthetic attributed and labeled graphs for benchmarking GNNs. Unlike earlier models such as ALBTER and GenCAT, which struggle to match the learnability characteristics of real datasets, SkyMap was specifically built to reproduce GNN performance with high fidelity across architectures. To achieve this, the model takes as input a set of graph metrics (such as covering class distributions, mixing patterns, and feature correlations), and uses them to guide a four-phase generation pipeline: generating subgraphs per class, combining them using a mixing matrix, and assigning node features via a class-feature matrix. This architecture allows SkyMap to capture both topological and semantic properties of real-world graphs, resulting in significantly lower Wasserstein distances in accuracy distributions between synthetic and original datasets. The model was evaluated using Wasserstein distance in order to show that Skymap is capable to recreate more accurately real-world graphs. 

Overall, existing attributed graph generators differ substantially in terms of flexibility, structural control, and applicability to graph learning tasks. While some methods focus on reproducing real-world structural properties or preserving degree distributions, others emphasize attribute-label relationships, heterophilic interactions, or data augmentation capabilities. However, most approaches do not jointly support attributed graph generation, graph cloning, augmentation, heterophily control, degree preservation, and explicit sub-community modeling. Table~\ref{tab:comparison_generators} summarizes the main characteristics of the discussed models and highlights the broader flexibility provided by SynCo in comparison with existing state-of-the-art approaches.

\begin{table*}[ht]
    \centering
    \caption{Comparison between different synthetic graph generation models}
    \resizebox{\textwidth}{!}{
     \begin{tabular}{|c|c|c|c|c|
        >{\centering\arraybackslash}p{3cm}|
        >{\centering\arraybackslash}p{2cm}|
        >{\centering\arraybackslash}p{3cm}|} 
        \hline 
        Model & Attributed  & Clone & Augment & Generator & Sub-Comm. Structure & Heterophily Control & Deg. Preservation/Control \\ 
        \hline
        DANCer &  &  & & \checkmark &  &  & \\ 
        Chung-Lu & & \checkmark &&  &  &  & \checkmark\\ 
        AL-BTER & \checkmark &  & &  \checkmark &  & \checkmark & \checkmark \\ 
        SkyMap & \checkmark & \checkmark  &  \checkmark&\checkmark &  & \checkmark & \checkmark \\ 
        GenCAT & \checkmark & \checkmark & \checkmark & \checkmark & & \checkmark & \\ 
        SynCo (Ours) &\checkmark  & \checkmark & \checkmark & \checkmark & \checkmark & \checkmark & \checkmark \\ 
        \hline
    \end{tabular}
    }
    \label{tab:comparison_generators}
\end{table*}

\subsection{Why Sub-Community Structure in Synthetic Graphs}

Synthetic graph generators designed for benchmarking or graph mimicking commonly preserve global or community-level properties, such as node degree distributions and intra- and inter-community connectivity patterns \cite{bonifati2021graph}. Although these properties are essential, they may not be sufficient to reproduce the internal organization of real communities. In many real-world networks, communities exhibit a hierarchical structure, in which smaller cohesive groups are embedded within larger communities \cite{Lancichinetti2009detecting}. Therefore, two graphs may present similar degree distributions and comparable community mixing patterns while differing substantially in their local organization.

\begin{figure*}[H]
    \centering
    \includegraphics[width=\linewidth]{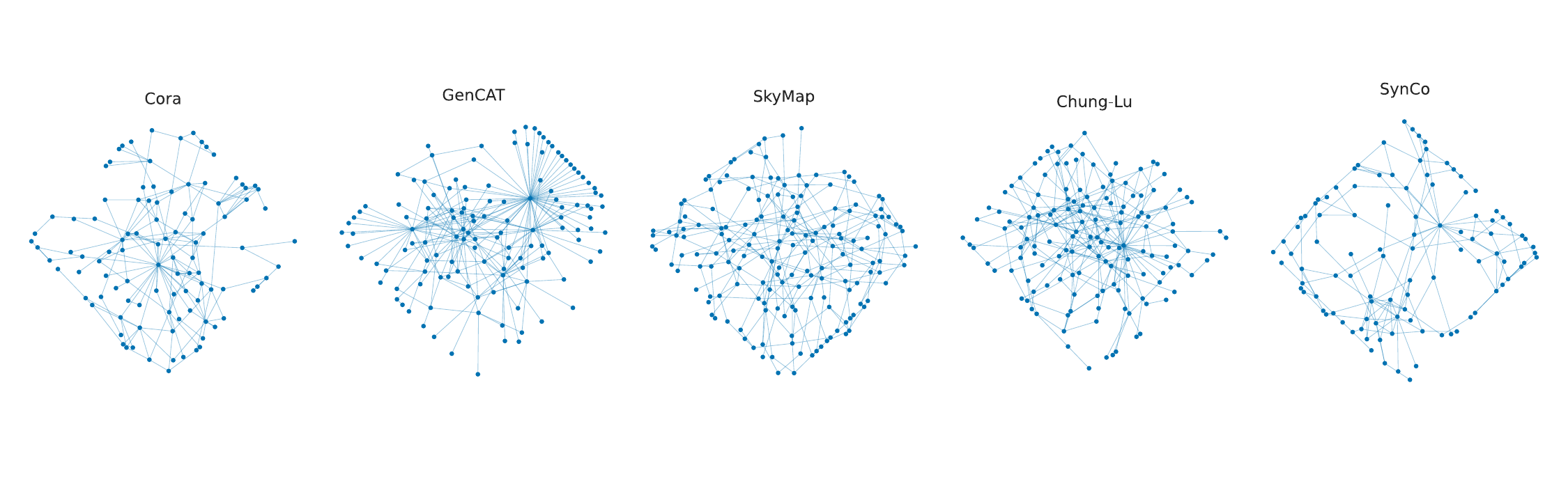}
    \caption{Mimicked Graph of each model over class label 5 of Cora dataset.}
    \label{fig:why_SCAtt}
\end{figure*}

This limitation is illustrated in Fig.~\ref{fig:why_SCAtt}, which presents the subgraph induced by a selected community from the Cora dataset and the corresponding structures generated by different graph generators. From a qualitative perspective, the original community contains an internally organized topology with localized dense regions and characteristic connectivity patterns. While conventional generators may reproduce some global properties of the graph, they can considerably reorganize the internal structure of the selected community. In contrast, SynCo is designed to retain this finer-grained organization by explicitly modeling relationships among sub-communities.

The incorporation of sub-community structure is therefore relevant in two complementary scenarios. In graph mimicking, it enables synthetic replicas to preserve not only global statistics and community membership patterns, but also the internal organization observed within each community. In controlled graph generation, sub-community parameters allow the construction of clusters with heterogeneous internal structures, enabling more challenging and realistic benchmarks for community detection and attributed graph clustering methods. This capability is particularly important under heavy-tailed degree distributions, such as power-law, where a small number of highly connected nodes can strongly influence the topology of an entire community.

\section{Synthetic Community-Aware Attributed Graph Generator}

To address datasets exhibiting heterogeneous densities, enable the control of specific node degree distributions, and incorporate different levels of homophily and heterophily in graphs with sub-community structures, we propose the \textit{Synthetic Community-Aware Attributed Graph Generator for Graph Neural Network Benchmarking} (SynCo). SynCo framework is designed to support the three tasks introduced in Section \ref{sec:problem_statement}, namely:  
(1) the generation of a fully synthetic graph based solely on user-defined parameters;  
(2) the cloning of an existing graph while preserving its structural properties; and  
(3) the augmentation of a cloned graph through controlled modifications.

Regardless of the selected task, the SynCo algorithm follows a unified pipeline composed of four sequential stages:  
(i) node community and sub-community assignment,  
(ii) edge construction,  
(iii) topological noise generation, and  
(iv) attribute assignment.  
Each stage plays a distinct role in ensuring that the resulting graph faithfully reflects both the community structure and the desired topological and attribute-related properties. The details of each stage are described in the following subsections.

\subsection{SynCo generation from user-defined inputs}

In the user-driven generation mode, the graph is fully specified through a collection of matrices, constants, and probability distributions provided as input. These parameters collectively describe the target community structure, intra- and inter-community connectivity patterns, and node degree characteristics.  
Table~\ref{tab:parameters} summarizes the complete set of inputs required for the construction of the output graph using the SynCo framework.

\begin{table*}[ht]
    \centering
    \caption{Input parameters of the SynCo graph generation algorithm.}
    \begin{tabular}{ccp{11cm}}
        \hline
        \textbf{Parameter} & \textbf{Type} & \textbf{Description} \\
        \hline
        $n$ & Integer & Total number of nodes in the graph. \\
        $k$ & Integer & Number of communities in the graph. \\
        $k_\ell$ & Integer & Number of sub-communities of community $C_\ell$ \\         $d$ & Integer & Dimension of feature matrix $X$ \\ 
        $\mathbf{y}$ & Vector & Vector of positive integers specifying the number of nodes in each community, with $\dim(\mathbf{y}) = k$. \\
        $\mathbf{e}$ & Vector & Vector of positive integers representing the number of homogeneous edges in each community, with $\dim(\mathbf{e}) = k$. \\
        
        $\mathbf{dst}$ & Set & Collection of node degree distributions vectors, where each element defines the node degree distribution associated with a community, with $\dim(\mathbf{dst}) = k$. \\
        $\mathcal{S}$ & Set & Collection of vectors $\mathcal{S} = \{ S_1, S_2, \cdots, S_k\}  | S_i \in \mathbb{R}^{k_i}$, where each vector specifies the number of nodes assigned to each sub-community within a community. \\
        $\rho$ & Integer & Total number of edges, i.e, heterogeneous (between communities) and homogeneous (within communities). \\
        $\mathcal{A}_{\text{in}}$ & Set & Collection of matrices $\mathcal{A}_{\text{in}} = \{ \mathcal{A}_{\text{in}}^{(1)}, \mathcal{A}_{\text{in}}^{(2)}, \cdots, \mathcal{A}_{\text{in}}^{(k)}\} | \mathcal{A}_{\text{in}}^{(i)} \in \mathbb{R}^{k_i \times k_i}$ defining the connection probabilities between sub-communities within each community. \\
        $\mathcal{A}_{\text{out}}$ & Matrix & Matrix encoding the number of edges allocated between communities, where $(\mathcal{A}_{\text{out}})_{i,i} = 0$. \\
        $\alpha_{\text{feat}}$ & Real & Noise parameter for feature generation in Equation \ref{eq:stochastic_attribute_initialization}\\ 
        $\alpha_{\text{topo}}$ & Real & Noise parameter for topological-feature aggregation in Equation \ref{eq:topological_noise} \\ \hline
    \end{tabular}
    \label{tab:parameters}
\end{table*}

\subsubsection{Node community and sub-community assignment stage}

Let $\mathbf{y} = \left[ y_1, y_2, \ldots, y_k \right]$ denote the vector specifying the number of nodes assigned to each of the $k$ communities, where $y_c$ corresponds to the cardinality of community $c$. Based on this information, the algorithm constructs a community label vector
$ \mathbf{Y} \in \mathbb{R}^{n},$
with $n = \sum_{c=1}^{k} y_c$, such that each entry $Y_i$ indicates the community label associated with node $i$. This procedure ensures that every node is deterministically assigned to exactly one community, fully respecting the predefined community sizes.

In addition to community-level assignments, SynCo introduces a finer-grained structure by defining sub-communities within each community. For a given community $c$, the vector $\mathcal{S}_c$ specifies the distribution of its sub-communities; that is, the $i$-th entry of $\mathcal{S}_c$ represents the probability that a node in community $c$ belongs to sub-community $i$. Based on this distribution, the algorithm assigns each node in community $c$ to exactly one of its corresponding sub-communities.

At the end of this stage, the algorithm possesses two fundamental pieces of information required for subsequent steps: each node is uniquely associated with both a community and a sub-community. This joint assignment provides the structural foundation for edge generation, enabling the controlled modeling of intra and inter sub-community connectivity patterns in the following stages.

\subsubsection{Edge Construction Stage}

In the edge construction stage, each community is assigned a predefined number of {homogeneous edges}, i.e., edges connecting nodes belonging to the same community, as specified by the vector $\mathbf{e}$, where $e_i$ denotes the number of homogeneous edges of community $i$. This design choice enables explicit control over the intra-community connectivity, making the resulting graph particularly suitable for community detection tasks.

In addition, each community is associated with a degree-generating distribution that governs the node degree distribution within that community. By controlling this distribution, the algorithm allows the formation of topologies with distinct structural roles, such as core and border nodes. The model supports  three degree distributions: \emph{normal}, \emph{power-law}, or \emph{uniform}.

For the normal distribution, the mean $\mu$ and standard deviation $\sigma$ are defined by default as
\begin{equation}
\mu = \frac{n + 1}{2}, \quad 
\sigma = \max\left(\frac{n}{6}, 1.0\right),
\end{equation}
where $n$ denotes the number of nodes in the corresponding sub-community. The probability density function is given by
\begin{equation}
f(x) = \frac{1}{\sqrt{2\pi\sigma^2}} \exp\left(-\frac{(x - \mu)^2}{2\sigma^2}\right).
\end{equation}

For the power-law distribution, the degree probability follows
\begin{equation}
f(x) = x^{-\alpha},
\end{equation}
with the default exponent set to $\alpha = 2$, enabling the emergence of hub-dominated structures.

For the uniform distribution, all nodes are sampled with equal probability:
\begin{equation}
    f(x) = \frac{1}{n}.
\end{equation}

For a given community $c$, the source and target sub-communities used to generate an edge are sampled according to the sub-community interaction matrix $\mathcal{A}_{\text{in}}$. Subsequently, the selection of the source and target nodes within their respective partitions is guided by the community-specific degree distribution $\mathbf{dst}_c$. Since these distributions are constructed independently for each partition, the algorithm preserves local structural properties while maintaining global consistency. To ensure that all nodes remain eligible for selection, a small constant value of $1 \times 10^{-6}$ is added to each probability.

This strategy enables the generation of multiple sub-communities sharing the same degree distribution while emphasizing local topological characteristics, such as centrality and connectivity heterogeneity.

\subsubsection{Noise Generation Stage}

At this stage, the graph consists exclusively of homogeneous edges and may remain disconnected. To reflect realistic network scenarios and to support node clustering benchmarks, the algorithm introduces structural noise in the form of {heterogeneous edges}, i.e., edges connecting nodes from different communities. The amount of noise is controlled through the total number of edges $\rho$, which determines the number of heterogeneous and homogeneous edges to be added based on the community interaction matrix $\mathcal{A}_\text{out}$, where $(\mathcal{A}_\text{out})_{i,j}$ represents the probability of connection between the communities $i$ and $j$.

This design provides the user with two complementary mechanisms to control the graph topology: (1) by specifying the internal structure through community sizes, sub-communities, and degree distributions; and (2) by adjusting the global number of edges $\rho$, which regulates the level of inter-community connectivity. Together, these mechanisms allow fine-grained control over both intra-community organization and inter-community heterogeneity.

\subsubsection{Attribute Generation Stage}

After the structural construction of the graph, node attributes are generated with a community-conditional and topology-aware behavior. This stage aims to produce feature representations that simultaneously preserve community separability and incorporate structural information derived from the graph topology.

Let ${G} = ({V}, {E})$ be the graph obtained after the edge construction and noise generation stages. The attribute matrix ${X} \in \mathbb{R}^{n\times d}$ is generated through a three-step process: (i) construction of community prototype vectors, (ii) injection of stochastic variability, and (iii) topology-aware feature smoothing.

\paragraph{Community-Conditional Prototype Construction}

Initially, each community is associated with a latent prototype vector in the attribute space. To ensure linear independence and avoid redundant feature directions across communities, an orthonormal basis is constructed via QR decomposition. Specifically, a random matrix
\[
\mathbf{B} \in \mathbb{R}^{d \times k}
\]
is sampled from a standard normal distribution (where $k \leq d$) , and its QR decomposition yields an orthonormal matrix $\mathbf{Q}$. The resulting community prototypes are defined as the rows of $\mathbf{Q}^\top$, such that each community $c$ is assigned a prototype vector $\mathbf{q}_c \in \mathbb{R}^d$.

\paragraph{Stochastic Attribute Initialization}
For each node $v \in {V}$ with community label ${y}_v = c$, the initial attribute vector $\mathbf{x}_v^{(0)}$ is generated as
\begin{equation}
\label{eq:stochastic_attribute_initialization}
    \mathbf{x}_v^{(0)} = \mathbf{q}_c + \alpha_\text{feat} \cdot \boldsymbol{\epsilon}_v,
\end{equation}
where $\boldsymbol{\epsilon}_v \sim \mathcal{N}(\mathbf{0}, \mathbf{I}_d)$ is a Gaussian noise vector and $\alpha_\text{feat} \in [0,1]$ controls the intra-community dispersion. This formulation ensures that nodes belonging to the same community are centered around a common prototype while maintaining sufficient variability to avoid degenerate feature distributions.

\paragraph{Degree-Normalized Topological-Aware Feature Smoothing}
To incorporate structural information into the attribute space, a single-step propagation mechanism inspired by graph signal processing is applied. Let $\mathbf{A}$ denote the adjacency matrix of $\tilde{G}$ and let $\mathbf{D}$ be a diagonal matrix defined as
\begin{equation}
\mathbf{D}_{vv} =
\begin{cases}
\deg(v)^{-1/2}, & \text{if } \deg(v) > 0, \\
0, & \text{otherwise}.
\end{cases}
\end{equation}
The final attribute matrix $\tilde{X}$ is obtained via the convex combination
\begin{equation}
\label{eq:topological_noise}
   \tilde{X} = \alpha_\text{topo} \tilde{X}^{(0)} + (1 - \alpha_\text{topo})\mathbf{D}\mathbf{A}\tilde{X}^{(0)}, 
\end{equation}
where $\alpha_\text{topo} \in [0,1]$ controls the trade-off between community-conditional feature purity and neighborhood-induced smoothing.

The proposed formulation provides explicit control over the statistical and structural properties of the generated attributes through the parameters $\alpha_\text{feat}$ and $\alpha_\text{topo}$. The parameter $\alpha_\text{feat}$ regulates the variance of the Gaussian perturbation applied during the attribute initialization, directly controlling the intra-community dispersion in the feature space. Smaller values of $\alpha_\text{feat}$ yield more compact and well-separated clusters, while larger values increase overlap between communities in the feature space, allowing the generation of more challenging scenarios for downstream learning tasks. 

Conversely, the parameter $\alpha_\text{topo}$ governs the relative contribution of community conditional information and topological smoothing in the final attribute representation. Higher values of $\alpha$ emphasize community prototypes and preserve feature purity, whereas lower values increase the influence of neighborhood aggregation, strengthening the coupling between node attributes and graph structure. Together, these parameters enable fine-grained adjustment of the attribute generation process, allowing the user to tailor the synthetic data to different levels of structural dependency and classification difficulty. Algorithm \ref{alg:SCAtt} describes the workflow of the proposed algorithm.

\begin{algorithm*}[t]
\caption{SynCo over user-inputs}
\label{alg:SCAtt}
\begin{algorithmic}
\Require Number of nodes $n$; community sizes $\mathbf{y}$; number of communities $k$;
homogeneous edge counts $\mathbf{e}$;
sub-community sizes $\mathcal{S}=\{S_1,\dots,S_k\}$;
intra-community interaction matrices $\mathcal{A}_{\mathrm{in}}=\{
\mathcal{A}_{\mathrm{in}}^{(1)},\dots,\mathcal{A}_{\mathrm{in}}^{(k)}\}$;
degree distribution types $\mathbf{dst}$;
total number of edges $\rho$;
inter-community edge allocation matrix $\mathcal{A}_{\mathrm{out}}$;
attribute parameters $(d,\alpha_{\mathrm{feat}},\alpha_{\mathrm{topo}})$.
\Ensure Attributed graph $\tilde{G}=(\tilde{V},\tilde{E},\tilde{X},\tilde{Y})$.

\State Initialize graph $\tilde{G}$ with $n$ nodes and empty edge set
\State Assign community labels $\tilde{Y}$ according to $\mathbf{y}$
\State Assign nodes of each community $c$ to sub-communities according to $S_c$

\vspace{2pt}
\Statex \textbf{Homogeneous Edge Construction}
\For{each community $c = 1,\dots,k$}
    \State Compute node sampling weights inside each sub-community using $dst_c$
    \While{the number of homogeneous edges in $c$ is smaller than $e_c$}
        \State Sample a source sub-community according to $S_c$
        \State Sample a target sub-community according to the corresponding row of $\mathcal{A}_{\mathrm{in}}^{(c)}$
        \State Sample nodes $u$ and $v$ from the selected sub-communities using their degree-distribution weights
        \If{$u \neq v$ and $(u,v) \notin \tilde{E}$}
            \State Add edge $(u,v)$ to $\tilde{E}$
        \EndIf
    \EndWhile
\EndFor

\vspace{2pt}
\Statex \textbf{Heterogeneous Edge Construction}
\State Set the target number of heterogeneous edges to $\rho - \sum_{c=1}^{k} e_c$
\While{the target number of heterogeneous edges has not been reached}
    \State Sample a source community according to row sums of $\mathcal{A}_{\mathrm{out}}$
    \State Sample a target community according to the selected row of $\mathcal{A}_{\mathrm{out}}$
    \State Sample nodes $u$ and $v$ from the selected communities
    \If{$u \neq v$, $\tilde{Y}_u \neq \tilde{Y}_v$, and $(u,v) \notin \tilde{E}$}
        \State Add edge $(u,v)$ to $\tilde{E}$
    \EndIf
\EndWhile

\vspace{2pt}
\Statex \textbf{Attribute Generation}
\State Generate orthogonal community prototype vectors in $\mathbb{R}^d$
\State Initialize node attributes with community prototypes and Gaussian perturbation controlled by $\alpha_{\mathrm{feat}}$
\State Apply degree-normalized topological smoothing controlled by $\alpha_{\mathrm{topo}}$
\State Compact node identifiers and return the attributed graph

\State \Return $\tilde{G}$
\end{algorithmic}
\end{algorithm*}

\subsection{SynCo for graph mimicking and augmentation}

Let $G = (V,E,X,Y)$ be an attributed graph, where $V$ is the set of nodes, $E$ is the set of edges, $X \in \mathbb{R}^{|V|\times d}$ the node attributed matrix, and $Y \in \{1, \cdots, C\}^{|V|}$ the community labels associated with each node. Additionally, assume that $G$ can be decomposed into community-induced subgraphs $G_c = (V_c, E_c)$, where $V_c = \{v \in V: y_v = c \}$. Nodes in each community-induced subgraphs are homogeneous, i.e, the set of edges $E_c$ connect elements that belong to the same community $c$.

For the graph mimicking and augmentation task, the objective of the proposed algorithm is to generate a new attributed graph $\tilde{G} = (\tilde{V}, \tilde{E}, \tilde{X}, \tilde{Y})$ that preserves the structural, community-level and connectivity homogeneity patterns observed in $G$, while optionally allowing the graph to be augmented to a larger number of nodes, based on node degree distributions of the original data. 

The way that the task will be concluded is similar to the graph generation via user-inputs task. So, in next sections we describe the main differences between the tasks.

\subsubsection{Node commmunity Assignment and sub-community Discovery Stage}

A new undirected and unweighted graph $\tilde{G}$ is initialized with the same number of nodes as the original graph $G$. The node label vector $\tilde{Y}$ is directly inherited from $Y$, ensuring community consistency at the node level.

To capture the topological organization within each sub-community in $C_\ell$, a community detection algorithm, Parallel Louvain Method (PLM) by default, is applied to identify a set of sub-communities
\[
\mathcal{P}_\ell = \{P_{\ell,1}, \ldots, P_{\ell,k_\ell}\}.
\]
Based on this partitioning, a sub-community interaction matrix $\mathcal{A_\text{in}}^{(\ell)} \in \mathbb{R}^{k_\ell \times k_\ell}$ is computed for each community, where each entry quantifies the empirical tendency of nodes in sub-community $P_{\ell,i}$ to establish connections with nodes in sub-community $P_{\ell,j}$.

\subsubsection{Edge and Noise Construction Stages}

After identifying the partitions associated with each community, the edge construction stage begins. The source and target partitions for every edge in the synthetic graph $\tilde{G}$ is guided by the relative sizes of the corresponding partitions in the original graph $G$. Subsequently, the selection of individual nodes within the chosen partitions is performed according to the degree distribution observed in $G$. This strategy ensures that both local and global topological characteristics are preserved in the replicated graph structure.

At this step of the algorithm, the graph is composed only of homogeneous edges. From $G$ it is possible to extract a matrix $\mathcal{A}_{\text{out}}$, where $(\mathcal{A}_\text{out})_{i,j}$ is the probability of a heterogeneous edge connect the communities $i$ and $j$. Thus, heterogeneous edges are added to $\tilde{G}$ until the target number of edges ${\rho}$ of $G$ is reached on $\tilde{G}$.

At this point of the algorithm, the topological structure is complete and it is based on the original graph. The absence of external distribution as in GenCAT \cite{maekawa2023gencat} and SkyMAP \cite{Wassington2024skymap} allow the generated graph to be faithful to the original one.

\subsubsection{Attribute Assignment Stage}

This stage aims to generate the attribute matrix $\tilde{X}$ of the mimicked graph $\tilde{G}$ by directly leveraging the node features of the original graph $G$. The underlying assumption is that nodes occupying similar structural roles within the same community should exhibit comparable attribute representations. To this end, a rank-based matching strategy is employed to transfer node attributes from $G$ to $\tilde{G}$ in a label-conditional and topology-aware manner.

The attribute generation process is performed independently for each community label $c \in \{1, \dots, k\}$. For a given community $c$, we first identify the subset of nodes $V_c \subset V$ such that $y_v = c$, and consider the corresponding nodes $\tilde{V}_c \subset \tilde{V}$ in the mimicked graph.

Within each community, nodes in both graphs are ordered according to their degree, resulting in two degree-ranked sequences: one derived from the original graph $G$ and another from the mimicked graph $\tilde{G}$. A one-to-one correspondence between nodes is then established by matching nodes with identical rank positions in these ordered lists. In this way, nodes with similar degree-based structural importance in $G$ are aligned with nodes playing analogous roles in $\tilde{G}$.

Once this correspondence is defined, the attribute vector of each node in the mimicked graph is initialized by copying the attribute vector of its matched counterpart in the original graph. Formally, for a matched node pair $(v, \tilde{v})$, the attribute assignment is given by
% To avoid exact duplication and to introduce controlled variability, an additive noise term sampled from a Gaussian distribution is applied to each transferred feature vector. 
\begin{equation}
\tilde{\mathbf{x}}_{\tilde{v}} = \mathbf{x}_{v} 
% + \boldsymbol{\epsilon},
\end{equation}

This rank-based attribute matching strategy preserves the empirical distribution of node features within each community while ensuring consistency with the reconstructed topology. By conditioning the transfer on both community membership and degree-based structural roles, the method maintains semantic coherence between structure and attributes, making the resulting synthetic graph suitable for downstream tasks such as clustering, node classification and graph representation learning in general.

\subsection{Graph Augmentation Stage}

In addition to mimicking an existing attributed graph, the proposed framework supports \emph{data augmentation} by increasing the size of the synthetic graph to a user-defined number of nodes. Given a target size $|\tilde{V}| = n'$, with $n' > |V|$, the algorithm adds $n' - |V|$ new nodes to the mimicked graph $\tilde{G}$ while preserving (i) the empirical community proportions, (ii) the sub-community structure within each label, and (iii) the degree-driven connectivity patterns observed in the reconstructed topology.

\paragraph{Partition model for the augmented graph}
Before inserting new nodes, the algorithm recomputes the partition structure within each community induced-subgraph of $\tilde{G}$ using the same community detector employed in the mimic stage. For each community $c$, this yields a set of sub-communities $\mathcal{P}_\ell$ and a corresponding sub-community interaction matrix $\mathcal{A}_{\text{out}}$, which summarizes the empirical tendency of edges to occur between sub-communities. This partition model is then used as a guide for placing and wiring new nodes in a topology-consistent way.

\paragraph{Community and sub-community assignment}
Each new node $v_i$ is first assigned to a community $c_i$ by sampling from the empirical community distribution observed in $\tilde{Y}$, i.e., communities with larger support in the mimicked graph are more likely to receive additional nodes. Conditioned on the sampled community, the algorithm assigns $v_i$ to a source partition $P_{\ell,s}$ by sampling from the partition probability distribution of that community (which is proportional to partition sizes). This step ensures that the augmented graph preserves both the global communities balance and the relative sizes of partitions inside each community.

\paragraph{Attribute initialization for new nodes}
After assign the node to a sub-community, the attribute vector of the new node is initialized to be representative of the local attribute distribution of its sub-community. Concretely, the feature vector $\tilde{\mathbf{x}}_{v_i}$ is set to the mean of the feature vectors of nodes currently belonging to $P_{\ell,s}$:
\[
\tilde{\mathbf{x}}_{v_i} = \frac{1}{|P_{\ell,s}|}\sum_{v \in P_{\ell,s}} \tilde{\mathbf{x}}_{v}.
\]
This strategy preserves community-level attribute coherence, ensuring that added nodes remain compatible with the attribute geometry learned during the mimic stage.

\paragraph{Edge formation via sub-community interactions}
To incorporate the new node into the graph structure, the algorithm creates a number of edges that reflects the typical connectivity of the selected partition. Specifically, the number of connections for $v_i$ is set to the average degree of homogeneous nodes in the sub-community $P_{\ell,s}$, approximated by
\[
m_i \approx \frac{2|E(P_{\ell,s})|}{|P_{\ell,s}|},
\]
with $m_i \geq 1$ to avoid isolating the new node.

For each of the $m_i$ edges, a target sub-community $P_{\ell,t}$ is sampled according to the interaction profile of the source sub-community, given by $\mathcal{A}_\text{in}$. Finally, a target node $v_j \in P_{\ell,t}$ is sampled with probability proportional to its current degree, promoting preferential attachment and preserving the degree heterogeneity of the reconstructed graph. The edge $(v_i,v_j)$ is added to $\tilde{E}$, and the interaction matrix is updated to reflect the newly created inter-partition connection.

Overall, the augmentation stage extends the mimicked graph without disrupting its learned organization. By sampling communities and sub-community according to their empirical prevalence, initializing attributes from sub-community statistics, and wiring new nodes through sub-community level interaction, the procedure yields larger synthetic graphs that remain structurally and semantically consistent with the original network.

\section{Experiments and Experimental Setup}

% In this section, we describe the experimental configuration adopted to evaluate the proposed model. We detail the methodological steps designed to analyze the behavior of SCAtt and to assess its performance in comparison with state-of-the-art attributed graph generators. The evaluation includes controlled parametric analysis, clustering benchmarks, and structural property validation.

% \gm{Nessa seção apresentamos as configurações experimentais e os experimentos adotados para avaliar o modelo proposto. A abordagem metodológica se divide principalmente em duas frentes: (1) Avaliação dos hiperparâmetros do algoritmo SCAtt com o objetivo de entender quais suas aplicações e limitações de geração de grafos em cenários mais ou menos homofílicos e presença de ruído estrutural ou semântico; (2) Avaliação de algoritmos de node clustering da literatura por meio do SCAtt, observando limitações e pontos fortes por meio do desempenho desses algoritmos em cenários mais ou menos desafiadores.}

In this section, we present the experimental setup and the experiments conducted to evaluate the proposed model. The methodological approach is structured around three main directions: (1) Hyperparameter evaluation of the SynCo algorithm, aimed at understanding its applicability and limitations in generating graphs under scenarios with varying levels of homophily and the presence of structural or semantic noise; and (2) Assessment of node clustering algorithms from the literature using SynCo, analyzing their strengths and limitations based on their performance in more or less challenging scenarios; and (3) power-law evaluation for mimicked graphs.

\subsection{Ablation Study}
\label{sec:ablationstudy}

{To evaluate the influence of the SynCo control parameters on the 
separability of the generated node attributes, we conducted a grid-search 
experiment over $\alpha_{\text{topo}}$ and $\alpha_{\text{feat}}$. A synthetic 
attributed graph with $n = 120$ nodes and $k = 3$ communities of equal size 
was generated, with each node represented by a $60$-dimensional feature 
vector. The communities were configured with $e = [250, 200, 100]$ and different degree distribution 
types: power-law, normal, and uniform, respectively. }

% Aditional graph configurations can be found at LINK}

{The parameters $\alpha_{\text{topo}}$ and $\alpha_{\text{feat}}$ were 
independently varied from 0 to 1 with increments of 0.05, resulting in 
$21 \times 21 = 441$ parameter combinations. This procedure was repeated 
for five values of $\rho$, namely 
$\{550, 700, 900, 1100, 1200\}$. To provide a controlled comparison 
among parameter configurations, all graphs were generated using the same 
random seed.}

\begin{figure*}[!htpb]
    \centering
    \includegraphics[width=\linewidth]{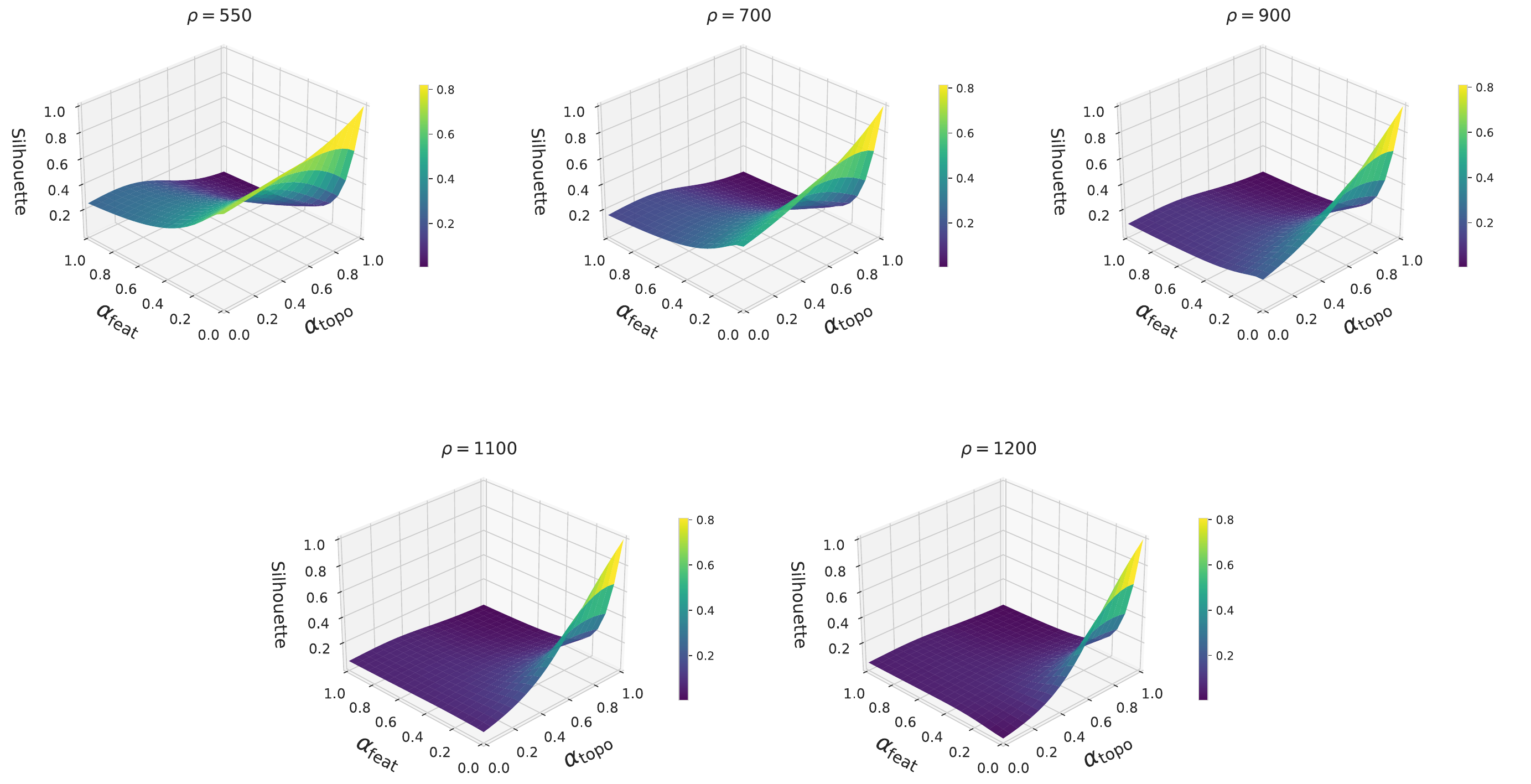}
    \caption{{Silhouette values for different generated graphs. As the total number of edges $\rho$ increase, the parameters $\alpha_\text{topo}$ and $\alpha_{\text{feat}}$ faster degrade the silhouette score.}}
    \label{fig:silhouette_SCAtt}
\end{figure*}

{For each generated graph, the node feature matrix and the corresponding 
ground-truth community labels were extracted. The quality of the separation 
in the attribute space was evaluated using the silhouette score, which 
measures the cohesion of nodes within the same community and their separation 
from nodes belonging to other communities. Since the score was calculated 
directly from the node attribute matrix, this experiment evaluates the effect 
of the SynCo parameters on feature separability. Figure \ref{fig:silhouette_SCAtt} illustrates the results of the experiment.}

\begin{figure*}[!htb]
    \centering
    \input{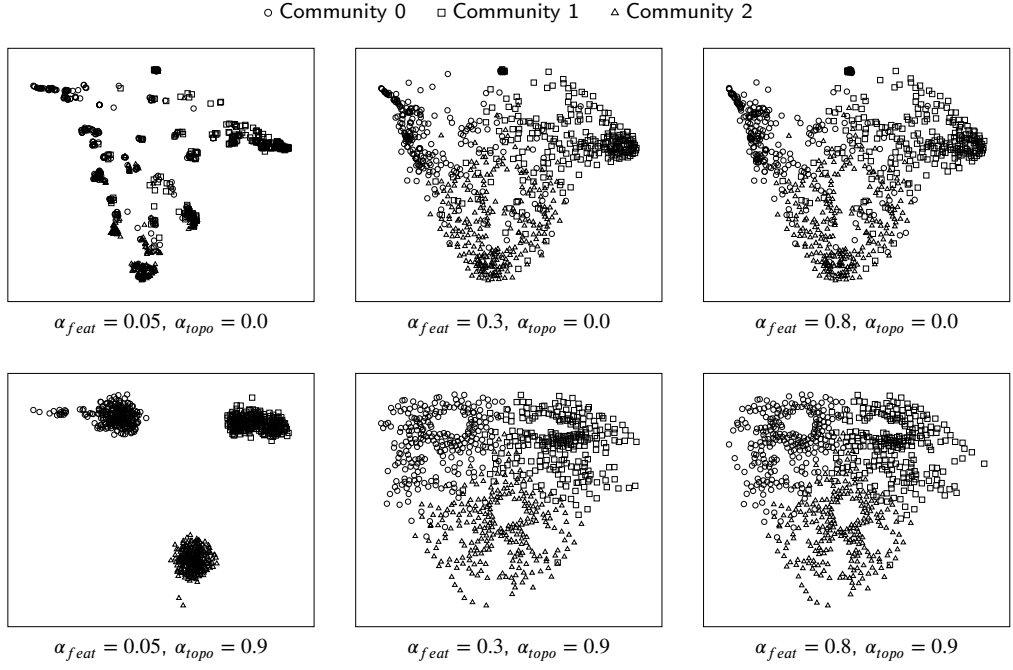}
    \caption{Feature space projection under different values of $\alpha_{feat}$ and $\alpha_{topo}$.}
    \label{fig:feature-space-alpha}
\end{figure*}

We also evaluate the embedding generated by different values of $\alpha_{\operatorname{feat}}$ and $\alpha_{\operatorname{topo}}$. Figure \ref{fig:feature-space-alpha} illustrates the t-SNE over the generated feature spaces.

% \gm{Nos grafos gerados com menor valor de $\rho$ os valores de $\alpha_\text{topo}$ and $\alpha_{\text{feat}}$ controlam melhor a região de separabilidade dos dados, principalmente nos intervalos $0.4 < \alpha_\text{topo}< 0.8$ e  $0 < \alpha_\text{topo}< 0.4$. Entretanto, quando valores de $\alpha_{\text{feat}}$ e $\alpha_{\text{topo}}$ se aproximam de 1, mesmo grafos completamente desconexos entre classes apresentam ruído semântico elevado. Para grafos com maior quantidade de arestas entre classe (ou seja, valor de $\rho$ maior quando comparado a quantidade de arestas homogêneas), os valores de $\alpha_{\text{feat}}$ e $\alpha_{\text{topo}}$ criam características separáveis em uma região muito maior. Isso mostra a maleabilidade do modelo proposto, sendo capaz de criar características separáveis em grupos menos densos, ou características ruidósas em grupos topológicos muito bem definidos.}

{For graphs generated with lower values of $\rho$, the parameters 
$\alpha_\text{topo}$ and $\alpha_\text{feat}$ provide finer control over 
the region in which the data remain separable, particularly within the ranges 
$0.4 < \alpha_\text{topo} < 0.8$ and $0 < \alpha_\text{feat} < 0.4$. 
However, as both $\alpha_\text{feat}$ and $\alpha_\text{topo}$ approach 1, 
even graphs whose communities are completely disconnected may exhibit high 
semantic noise. For graphs with a larger number of inter-community edges, that is, 
higher values of $\rho$ relative to the number of homophilous edges, 
$\alpha_\text{feat}$ and $\alpha_\text{topo}$ yield separable feature 
representations over a substantially larger region of the parameter space. These results demonstrate the flexibility of the proposed model, as it can 
generate separable feature representations even for sparsely connected groups, 
while also producing semantically noisy representations for groups with a 
highly well-defined topological structure.}

% Para avaliar o SCAtt no cenario topologico verificar a heterofilia/homofilia de alguma forma

\begin{figure*}
    \centering
    \includegraphics[width=\linewidth]{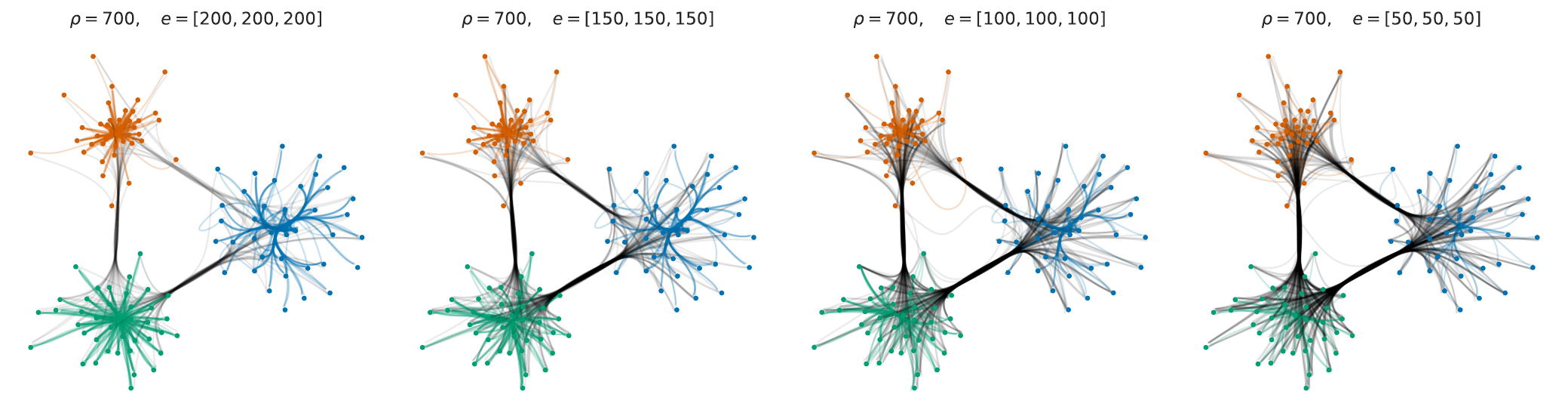}
    \caption{{Generated graphs with low heterophily (left) and high heterophily (right)}}
    \label{fig:heterogeneous_control}
\end{figure*}

{To illustrate the ability of the proposed model to generate graphs with 
different levels of homophily and heterophily, Fig.~\ref{fig:heterogeneous_control} 
shows four graphs generated under the same configuration, while varying the 
intra-community edge count vector $\mathbf{e}$ and keeping $\rho$ fixed. As the 
values of $\mathbf{e}$ decrease, the internal connectivity of each community 
becomes weaker, whereas inter-community connections become increasingly prominent. 
Consequently, the generated graphs progressively shift from a predominantly 
homophilous structure toward a more heterophilous configuration. This example 
highlights the capacity of the proposed generator to explicitly control the 
topological separability of the communities.}

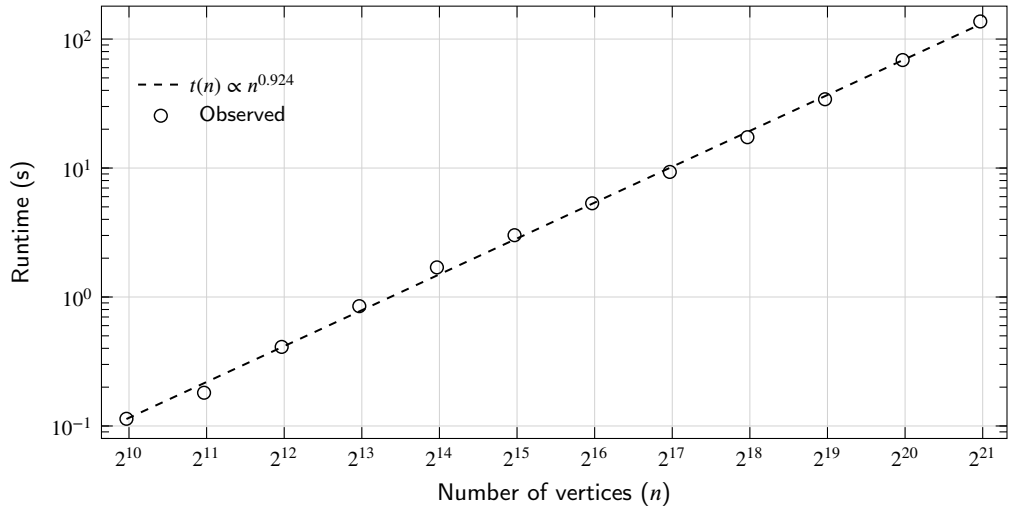
\begin{figure*}[!htb]
\centering
\begin{tikzpicture}
\begin{axis}[
    width=0.78\textwidth,
    height=0.42\textwidth,
    xmode=log,
    ymode=log,
    log basis x=2,
    log basis y=10,
    xlabel={Number of vertices ($n$)},
    ylabel={Runtime (s)},
    xmin=800,
    xmax=2600000,
    ymin=0.08,
    ymax=180,
    grid=major,
    grid style={line width=0.2pt, draw=gray!35},
    axis line style={black, line width=0.4pt},
    tick style={black, line width=0.4pt},
    legend style={
        draw=none,
        fill=none,
        font=\footnotesize,
        at={(0.03,0.87)},
        anchor=north west
    },
    tick label style={font=\footnotesize},
    label style={font=\small},
]

% Log-log fitted curve: log(t) = 0.924167 log(n) - 8.562548
\addplot+[
    no marks,
    black,
    dashed,
    line width=0.75pt
] coordinates {
    (1000,0.113197)
    (2000,0.214801)
    (4000,0.407603)
    (8000,0.773463)
    (16000,1.467715)
    (32000,2.785118)
    (64000,5.285007)
    (128000,10.028768)
    (256000,19.030472)
    (512000,36.112000)
    (1024000,68.525708)
    (2048000,130.033580)
};
\addlegendentry{$t(n)\propto n^{0.924}$}

% Observed runtimes
\addplot+[
    only marks,
    mark=o,
    mark size=2.4pt,
    black,
    line width=0.45pt,
    mark options={solid, fill=white, draw=black, line width=0.55pt}
] coordinates {
    (1000,0.113693)
    (2000,0.181041)
    (4000,0.410295)
    (8000,0.848891)
    (16000,1.696811)
    (32000,3.014987)
    (64000,5.331213)
    (128000,9.329173)
    (256000,17.340251)
    (512000,34.152654)
    (1024000,68.764787)
    (2048000,136.918474)
};
\addlegendentry{Observed}

\end{axis}
\end{tikzpicture}
\caption{Empirical runtime of SynCo as the number of vertices increases. The number of edges is fixed as $\rho = 5n$.}
\label{fig:synco-time-complexity}
\end{figure*}

The computational cost of SynCo is mainly determined by graph construction and attribute generation. Community and sub-community assignments require $\mathcal{O}(n)$ operations, while the homogeneous and heterogeneous edge construction stages require, in expectation, $\mathcal{O}(\rho)$ operations. Attribute initialization requires $\mathcal{O}(nd)$, and the topology-aware smoothing step requires $\mathcal{O}(\rho d)$ when the adjacency matrix is stored in sparse form. Therefore, ignoring the comparatively small cost associated with community interaction matrices, the overall time complexity of SynCo is $\mathcal{O}\bigl(d(n+\rho)\bigr)$. Figure \ref{fig:synco-time-complexity} summarize the execution time for different sizes of nodes and edges.

\subsection{Effective Evaluation in different scenarios}

To evaluate a range of node clustering algorithms under different scenarios, we use SynCo to generate graphs varying in homophily, number of nodes, number of communities, and community densities. To assess the strengths and weaknesses of these algorithms, we selected 10 approaches representing diverse taxonomies as classified by \cite{Messias2025systematic}. Table \ref{tab:nc_algorithms} summarizes the selected algorithms along with their corresponding classes.

\begin{table}[!htb]
    \centering
    \caption{Node clustering algorithms}
    \begin{tabular}{|c|c|} \hline
        \textbf{Algorithm} & \textbf{Classified Taxonomy}\\ \hline
        CoANE \cite{hsieh2023coane} & Reformulation Model\\
        CoANE\_topk \cite{hsieh2023coane} & Reformulation Model\\
        DGC-EFR \cite{Hao2022deep} & KL-Div Model\\
        SOLI \cite{molaei2021deep} & Topological Model\\
        DSAGC \cite{chen2023deep} & Reformulation Model\\
        DGCluster \cite{Bhowmick2024dgcluster}& Optimization Model\\
        NS4GC \cite{liu2024reliable} & Topological Model\\
        SANECE \cite{liao2021structure} & Reformulation Model\\
        UCoDe \cite{Moradan2023ucode}& Optimization Model\\
        EG-VGAE \cite{Cheng2024unveiling} & Optimization Model\\ \hline
    \end{tabular}
    
    \label{tab:nc_algorithms}
\end{table}

The tests were conducted to evaluate these models across three different aspects: (1) benchmarking the models in low- and high-heterophily scenarios; (2) benchmarking the models with a growing number of nodes per cluster; and (3) benchmarking the models when existing clusters are collapsed into smaller ones. The results are described below.

\paragraph{Collapsing Clusters Scenario}

To generate graphs in a cluster-collapsing scenario, SynCo produced seven distinct graphs. For the initial scenario, we generated a synthetic graph with $k=9$ communities, each containing 200 nodes. Each community was initialized with 500 internal edges, while the total number of edges in the graph was fixed at $\rho=5500$. The communities were organized into three groups of distribution patterns: power-law, normal, and uniform, with three communities per group. In the initial graph $G_0$, the matrix $A_{out}$ assigns uniform weight to connections between any pair of distinct communities, i.e., $(A_{out})_{ij}=1/8$ for $i \neq j$, with a zero diagonal. Node attributes were generated in dimension $d=60$, using $\alpha_{feat}=0.5$ and $\alpha_{topo}=0.5$, to equally balance the information contributed by features and topological structure.

Starting from the base scenario, the subsequent graphs were generated through a gradual reduction in the size and internal density of some communities. In each distribution group, one community was maintained with 200 nodes and 500 internal edges, while the other two progressively reduced in size and number of internal edges. Across graphs $G_1$ to $G_6$, these values decreased from 180 to 80 nodes and from 460 to 160 internal edges.

Table \ref{tab:comm_collapse_nmi} summarizes the results obtained by the algorithms under the community collapse scenario.

\begin{table*}[ht]
\centering
\caption{Model performance by collapse level in NMI.}
\label{tab:comm_collapse_nmi}
\begin{tabular}{lccccccc}
\toprule
Model & $G_0$ & $G_1$ & $G_2$ & $G_3$ & $G_4$ & $G_5$ & $G_6$ \\
\midrule
DGCluster & 0.907 & 0.858 & 0.840 & 0.772 & 0.737 & 0.750 & 0.688 \\
NS4GC & 0.894 & 0.834 & 0.815 & 0.792 & 0.761 & 0.691 & 0.533 \\
CoANE\_topk & 0.882 & 0.817 & 0.824 & 0.778 & 0.787 & 0.726 & 0.651 \\
CoANE & 0.865 & 0.757 & 0.783 & 0.735 & 0.728 & 0.718 & 0.630 \\
DSAGC & 0.847 & 0.754 & 0.710 & 0.709 & 0.667 & 0.645 & 0.603 \\
SANECE & 0.844 & 0.699 & 0.683 & 0.662 & 0.541 & 0.601 & 0.531 \\
SCGC & 0.801 & 0.704 & 0.691 & 0.618 & 0.640 & 0.681 & 0.592 \\
DGC & 0.783 & 0.740 & 0.716 & 0.667 & 0.648 & 0.698 & 0.578 \\
SOLI & 0.760 & 0.692 & 0.633 & 0.635 & 0.637 & 0.557 & 0.601 \\
EGC\_VGAE & 0.734 & 0.723 & 0.730 & 0.711 & 0.715 & 0.718 & 0.658 \\
UCODE & 0.091 & 0.108 & 0.090 & 0.044 & 0.084 & 0.131 & 0.223 \\
\bottomrule
\end{tabular}
\end{table*}

Overall, the results obtained by each algorithm indicate that community collapse tends to reduce the NMI achieved by the models. In almost all cases, the decrease is progressive, while in a small number of instances there is some fluctuation in intermediate scenarios, particularly in $G_3$ and $G_4$. This suggests that the collapse does not affect all methods linearly, especially for models such as CoANE, SCGC, and EGC\_VGAE.

\paragraph{Heterophily Increase Scenario}

\begin{figure*}[!htb]
    \centering

    \begin{tikzpicture}

        % Estilos das curvas
        \pgfplotsset{
            modelA/.style={
                blue,
                mark=*,
                mark size=1.8pt,
                line width=0.9pt
            },
            modelB/.style={
                red,
                mark=square*,
                mark size=1.7pt,
                line width=0.9pt,
                dashed
            },
            modelC/.style={
                teal!70!black,
                mark=triangle*,
                mark size=2.0pt,
                line width=0.9pt,
                dashdotted
            },
            modelD/.style={
                violet,
                mark=diamond*,
                mark size=1.9pt,
                line width=0.9pt,
                dotted
            }
        }

        \begin{groupplot}[
    group style={
        group size=1 by 4,
        vertical sep=1.25cm
    },
    width=15.0cm,
    height=4.1cm,
    xmin=0.08,
    xmax=0.92,
    ymin=0.00,
    ymax=0.90,
    xtick={0.1,0.2,0.3,0.4,0.5,0.6,0.7,0.8,0.9},
    ytick={0.0,0.2,0.4,0.6,0.8},
    grid=both,
    major grid style={gray!35},
    minor grid style={gray!15},
    minor tick num=1,
    tick label style={font=\scriptsize},
    label style={font=\small},
    title style={
        font=\small\bfseries,
        yshift=-4pt
    },
    ylabel={NMI},
    legend style={
        font=\scriptsize,
        draw=none,
        fill=white,
        fill opacity=0.85,
        text opacity=1,
        cells={anchor=west},
        /tikz/every even column/.append style={column sep=0.15cm}
    }
]

        % =====================================================
        % Painel 1: Reformulation Model
        % =====================================================
        \nextgroupplot[
            title={Reformulation Model},
            xticklabels=\empty,
            legend style={
                at={(0.98,0.97)},
                anchor=north east,
                legend columns=4
            }
        ]

        \addplot+[modelA] coordinates {
            (0.1,0.638343)
            (0.2,0.505190)
            (0.3,0.514043)
            (0.4,0.471013)
            (0.5,0.341426)
            (0.6,0.289152)
            (0.7,0.122019)
            (0.8,0.043757)
            (0.9,0.011030)
        };
        \addlegendentry{CoANE}

        \addplot+[modelB] coordinates {
            (0.1,0.719863)
            (0.2,0.682520)
            (0.3,0.656022)
            (0.4,0.651459)
            (0.5,0.566812)
            (0.6,0.490981)
            (0.7,0.370763)
            (0.8,0.275106)
            (0.9,0.077835)
        };
        \addlegendentry{CoANE top-k}

        \addplot+[modelC] coordinates {
            (0.1,0.661609)
            (0.2,0.610516)
            (0.3,0.514911)
            (0.4,0.318110)
            (0.5,0.220856)
            (0.6,0.113522)
            (0.7,0.053268)
            (0.8,0.024918)
            (0.9,0.011961)
        };
        \addlegendentry{DSAGC}

        \addplot+[modelD] coordinates {
            (0.1,0.768834)
            (0.2,0.615600)
            (0.3,0.520506)
            (0.4,0.421156)
            (0.5,0.271190)
            (0.6,0.108247)
            (0.7,0.045440)
            (0.8,0.024361)
            (0.9,0.019415)
        };
        \addlegendentry{SANECE}

        % =====================================================
        % Painel 2: KL-Div Model
        % =====================================================
        \nextgroupplot[
            title={KL-Div Model},
            xticklabels=\empty,
            legend style={
                at={(0.98,0.97)},
                anchor=north east,
                legend columns=1
            }
        ]

        \addplot+[modelA] coordinates {
            (0.1,0.806760)
            (0.2,0.792507)
            (0.3,0.731082)
            (0.4,0.667919)
            (0.5,0.569332)
            (0.6,0.549438)
            (0.7,0.375407)
            (0.8,0.356711)
            (0.9,0.139600)
        };
        \addlegendentry{DGC-EFR}

        % =====================================================
        % Painel 3: Topological Model
        % =====================================================
        \nextgroupplot[
            title={Topological Model},
            xticklabels=\empty,
            legend style={
                at={(0.98,0.97)},
                anchor=north east,
                legend columns=2
            }
        ]

        \addplot+[modelA] coordinates {
            (0.1,0.814060)
            (0.2,0.655307)
            (0.3,0.535393)
            (0.4,0.445110)
            (0.5,0.301339)
            (0.6,0.157972)
            (0.7,0.080473)
            (0.8,0.023085)
            (0.9,0.008988)
        };
        \addlegendentry{NS4GC}

        \addplot+[modelB] coordinates {
            (0.1,0.679674)
            (0.2,0.589778)
            (0.3,0.485761)
            (0.4,0.384810)
            (0.5,0.314069)
            (0.6,0.210797)
            (0.7,0.065236)
            (0.8,0.027278)
            (0.9,0.013295)
        };
        \addlegendentry{SOLI}

        % =====================================================
        % Painel 4: Optimization Model
        % =====================================================
        \nextgroupplot[
            title={Optimization Model},
            xlabel={Heterophily},
            legend style={
                at={(0.98,0.97)},
                anchor=north east,
                legend columns=3
            }
        ]

        \addplot+[modelA] coordinates {
            (0.1,0.830470)
            (0.2,0.656264)
            (0.3,0.485502)
            (0.4,0.382556)
            (0.5,0.251347)
            (0.6,0.142240)
            (0.7,0.061251)
            (0.8,0.035621)
            (0.9,0.026755)
        };
        \addlegendentry{DGCluster}

        \addplot+[modelB] coordinates {
            (0.1,0.737552)
            (0.2,0.641531)
            (0.3,0.561627)
            (0.4,0.475840)
            (0.5,0.362522)
            (0.6,0.244659)
            (0.7,0.138311)
            (0.8,0.067684)
            (0.9,0.028941)
        };
        \addlegendentry{EG-VGAE}

        \addplot+[modelC] coordinates {
            (0.1,0.580764)
            (0.2,0.461459)
            (0.3,0.341938)
            (0.4,0.265375)
            (0.5,0.090856)
            (0.6,0.040012)
            (0.7,0.022587)
            (0.8,0.020649)
            (0.9,0.010056)
        };
        \addlegendentry{UCoDe}

        \end{groupplot}

    \end{tikzpicture}

    \caption{NMI performance of graph clustering models under increasing heterophily levels, grouped by model taxonomy.}
    \label{fig:heterophily_nmi_models}
\end{figure*}
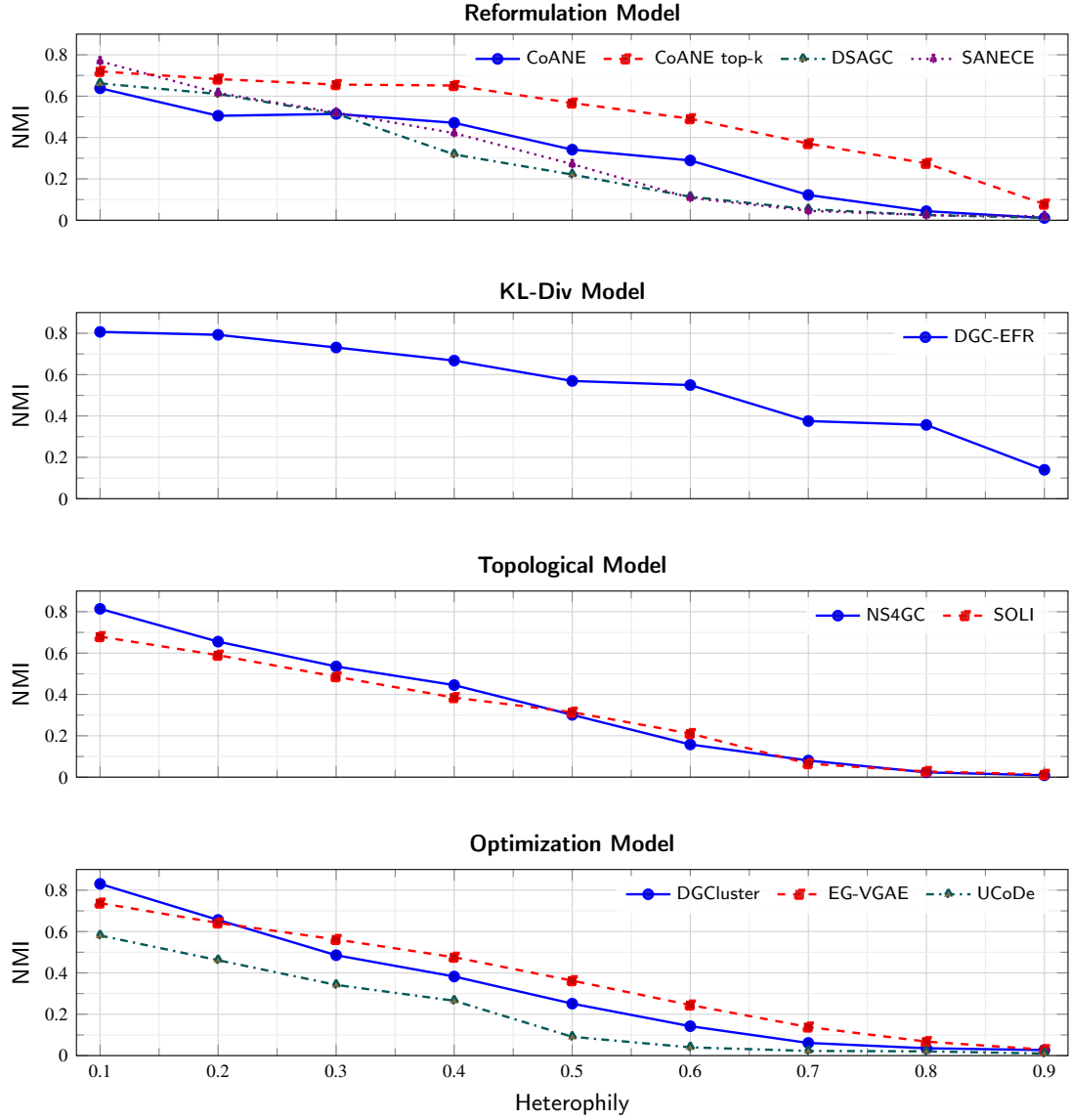

To evaluate the robustness of node clustering algorithms on graphs with increasing heterophily, we used SynCo to generate graphs with homophily rates $h$ ranging from 0.1 to 0.9, where edges are 10\% to 90\% heterogeneous, while keeping the total number of edges $\rho$ fixed. That is, for each graph generated with a higher $h$, the number of internal edges within each cluster decreases. Figure \ref{fig:heterophily_nmi_models} illustrates the results obtained in this scenario.

% ------------------------------------------------------------
% Figura
% ------------------------------------------------------------

Examining the results and comparing the models listed in Table~\ref{tab:nc_algorithms}, we observe a general trend of decreasing NMI values as the proportion of heterophilic edges increases. However, this decline is not uniform across all classes of models. Topology-based and optimization-based models tend to experience a steeper drop compared to KL-Divergence and embedding-based models. Since heterophily is directly related to topological characteristics, it is expected that optimization-based models, which rely on geometric network information, are less robust to increasing heterophily. Similarly, topology-based models, which focus on modifications to edge weights or network connections, inherently assume homophily to perform clustering effectively, making them more sensitive to heterophilic edges.

\paragraph{Graph Size Scenario}

To evaluate the performance of the algorithms as the number of nodes increases, we fixed the number of communities generated by SynCo at three while incrementally increasing the number of nodes $n$ for each generated graph. This setup allows us to assess both algorithmic performance and memory usage. Table~\ref{tab:graph_size_nmi} summarizes the main results obtained in this experimental stage. OOM stands for Out of Memory.

\begin{table*}[ht]
\centering
\caption{Model performance on graphs with varying numbers of nodes measured by NMI.}
\label{tab:graph_size_nmi}
\resizebox{\textwidth}{!}{%
\begin{tabular}{lcccccccccc}
\toprule
Model & $n=600$ & $n=1200$ & $n=1800$ & $n=2400$ & $n=3000$ & $n=3600$ & $n=4200$ & $n=4800$ & $n=5400$ & $n=6000$ \\
\midrule
DGC          & 0.633 & 0.792 & 0.773 & 0.702 & 0.766 & 0.742 & 0.759 & 0.738 & 0.752 & 0.731 \\
CoANE\_topk  & 0.695 & 0.675 & 0.693 & 0.690 & 0.711 & 0.699 & 0.695 & OOM   & OOM & OOM \\
SANECE       & 0.740 & 0.686 & 0.715 & 0.700 & 0.686 & 0.650 & 0.664 & 0.653 & 0.654 & 0.647 \\
NS4GC        & 0.688 & 0.667 & 0.684 & 0.683 & 0.675 & 0.664 & 0.666 & 0.673 & 0.667 & 0.657 \\
DGCluster    & 0.705 & 0.661 & 0.695 & 0.681 & 0.675 & 0.658 & 0.660 & 0.658 & 0.654 & 0.644 \\
EGC\_VGAE    & 0.643 & 0.641 & 0.652 & 0.647 & 0.636 & 0.626 & 0.647 & 0.635 & 0.644 & 0.603 \\
SOLI         & 0.586 & 0.599 & 0.564 & 0.574 & OOM   & OOM & OOM   & OOM   & OOM & OOM   \\
DSAGC        & 0.600 & 0.534 & 0.564 & 0.486 & 0.595 & 0.496 & 0.563 & 0.433 & 0.453 & 0.587 \\
CoANE        & 0.595 & 0.549 & 0.526 & 0.542 & 0.470 & 0.262 & 0.495 & 0.449 & 0.486 & 0.454 \\
SCGC         & 0.500 & 0.547 & 0.505 & 0.471 & 0.518 & OOM   & OOM & OOM & OOM   & OOM \\
UCODE        & 0.160 & 0.069 & 0.078 & 0.017 & 0.028 & 0.058 & 0.041 & 0.049 & 0.043 & 0.038 \\
\bottomrule
\end{tabular}%
}
\end{table*}

Overall, the results indicate that, in general, the models maintain consistent performance as the network size increases. This not only demonstrates the stability of these algorithms but also highlights the capability of the proposed model to generate networks of varying sizes while preserving well-defined topological and semantic characteristics.

\subsection{Scale-Free Analysis}

In order to evaluate the preservation of topological properties during the graph mimicking process, which involves data cloning and an increase in the number of nodes, we apply the scale-free plausibility test proposed by \cite{Clauset2009Power}. After assessing the plausibility of the power-law hypothesis, each network is classified according to its degree of adherence to this distribution, following an approach similar to that adopted in \cite{Broido2019scale}. The objective of this test is to evaluate whether the node degree distribution can be plausibly described by a power-law.

Let $G$ be a graph whose degree sequence $\{x_i\} = x_1, x_2, \cdots, x_k$ is modeled by a power-law distribution of the form
\[
\Pr(x) = Cx^{-\alpha}, \quad \alpha > 1, \quad x \geq x_{\min} \geq 1,
\]
where $\alpha$ denotes the scaling exponent, $C$ is the normalization constant, and $x$ represents the integer-valued node degrees. This modeling is applied exclusively to the upper tail of the distribution, defined by the set of values $x \geq x_{\min}$.

To identify the power-law distribution that best describes the degree distribution of graph $G$, it is necessary to determine an optimal value of $x_{\min}$ that marks the beginning of the upper tail, as well as to estimate the corresponding value of the exponent $\alpha$ using only the truncated data.

The procedure begins by evaluating each candidate value of $x_{\min}$. For each candidate, the exponent $\hat{\alpha}$ is estimated via maximum likelihood for the discrete distribution according to
\begin{equation}
    \hat{\alpha} = 1 + n \left[ \sum_{i=1}^n \ln\left(\frac{x_i}{x_{\min} - \frac{1}{2}}\right) \right]^{-1},
\end{equation}
where $x_i$, $i = 1, \cdots , n$, are the observed values such that $x_i \geq x_{\min}$. The estimated value of $\hat{\alpha}$ is then used to assess how well a power-law explains the tail of the distribution.

Let $S(x)$ denote the empirical cumulative distribution function (CDF) of the tail data, and let $P(x)$ denote the theoretical CDF of a power-law with parameters $\hat{\alpha}$ and $x_{\min}$. The selection of $x_{\min}$ is performed using the Kolmogorov-Smirnov statistic, defined as
\begin{equation}
\label{eq:Dvalue}
D(x_{\min}) = \max_{x \geq x_{\min}} \left| S(x) - P(x) \right|.
\end{equation}

The optimal value $\hat{x}_{\min}$ is obtained by minimizing the statistic $D(x_{\min})$, that is,
\[
\hat{x}_{\min} = \arg\min_{x_{\min}} D(x_{\min}).
\]

However, minimizing the Kolmogorov-Smirnov statistic alone does not guarantee that the degree distribution is adequately described by a power-law. To assess the plausibility of the fit, a goodness-of-fit test based on Monte Carlo simulations is applied. In this test, 1000 synthetic distributions are generated from power-law models whose parameters are estimated from the observed data.

For each synthetic replicate, the parameters $x_{\min}$ and $\hat{\alpha}$ are re-estimated, and the corresponding value of the statistic $D$ (Equation~\ref{eq:Dvalue}) is computed. The $p$-value is defined as the fraction of replicates for which the value of $D$ is greater than or equal to the value observed in the real data. Values of $p > 0.1$ indicate that the power-law hypothesis is plausible for the analyzed distribution.

Then, the distribution that better fits the data is compared to alternative distributions via likelihood test. 

In order to evaluate the different graphs generated over different numbers of edges, we use an approach similar to \cite{Broido2019scale}, where we create a ton of rules that guide the classification of every community in the graph that we want to evaluate the power-law plausibility. Each subgraph is classified as ``Not Scale-Free", ``Weak Scale-Free" and ``Strong Scale-Free". 

\begin{itemize}

    \item \textbf{Strong Scale-Free} regard over networks that have a $p$-value greater than 0.1, the estimated value of $\hat{\alpha}$ is between 2.0 and 3.0, and no alternative distribution is favored.

    \item \textbf{Weak Scale-Free} regard over networks that also have a $p$-value greater than 0.1, and no other alternative is favored, but the value of $\hat{\alpha}$ is not between the interval of $[2.0,3.0]$
    
    \item \textbf{Not Scale-Free} networks are the ones that do have the $p$-value lower than 0.1 or some alternative is favored.
\end{itemize}

We compare the proposed approach with GenCAT \cite{maekawa2023gencat} and Chung–Lu \cite{miller2011efficient} on the CiteSeer and PubMed datasets. Tables \ref{tab:sfpubmed} and \ref{tab:sfciteseer} present the classification results for the cloned datasets at their original size and when augmented by 1.3× and 1.5×. Subgraphs highlighted in bold indicate those classified into the same category as the original dataset.

% PubMed
\begin{table*}[ht]
\centering
\caption{Scale-Free Classification of PubMed dataset.}
\resizebox{\textwidth}{!}{
\begin{tabular}{|ccc|ccc|ccc|ccc|c|c|}
\hline
\multicolumn{3}{|c}{$\alpha$} & \multicolumn{3}{|c|}{$n_\text{tail}$} & \multicolumn{3}{c|}{$D$} & \multicolumn{3}{c|}{Class} & Subgraph & Model \\
$1 \times$ & $1.3\times$ & $1.5\times$ & $1 \times$ & $1.3\times$ & $1.5\times$ & $1 \times$ & $1.3\times$ & $1.5\times$ & $1 \times$ & $1.3\times$ & $1.5\times$ &  &  \\
\hline
2.19 & 2.11 & 2.13 & 247 & 547 & 651 & 0.02 & 0.03 & 0.03 & \textbf{Strong} & \textbf{Strong} & \textbf{Strong} & 0 & GenCAT \\
2.27 & 2.10 & 2.00 & 1058 & 1758 & 3609 & 0.02 & 0.01 & 0.02 & {Strong} & Strong & Strong & 1 & GenCAT \\
2.26 & 2.08 & 2.01 & 739 & 1754 & 4051 & 0.01 & 0.01 & 0.02 & {Strong} & Strong & Strong & 2 & GenCAT \\ \hline
2.21 & - & - & 1187 & - & - & 0.05 & - & - & \textbf{Strong} & - & - & 0 & chung-lu \\
3.14 & - & - & 401 & - & - & 0.04 & - & - & \textbf{Weak} & - & - & 1 & chung-lu \\
4.14 & - & - & 233 & - & - & 0.03 & - & - & \textbf{Weak} & - & - & 2 & chung-lu \\ \hline
2.22 & 2.56 & 2.61 & 1215 & 1529 & 1961 & 0.05 & 0.02 & 0.02 & \textbf{Strong} & \textbf{Strong} & \textbf{Strong} & 0 & SynCo \\
3.21 & 2.39 & 2.55 & 398 & 3925 & 3675 & 0.03 & 0.02 & 0.02 & \textbf{Weak} & {Strong} & Strong & 1 & SynCo  \\
4.17 & 4.29 & 2.42 & 195 & 161 & 4021 & 0.03 & 0.04 & 0.03 & \textbf{Weak} & \textbf{Weak} & Strong & 2 & SynCo  \\ \hline
2.03 & - & - & 1629 & - & - & 0.05 & - & - & Strong & - & - & 0 & PubMed \\
3.11 & - & - & 455 & - & - & 0.03 & - & - & Weak & - & - & 1 & PubMed \\
4.22 & - & - & 137 & - & - & 0.03 & - & - & Weak & - & - & 2 & PubMed \\
\hline
\end{tabular}
}

\label{tab:sfpubmed}
\end{table*}

% CiteSeer
\begin{table*}[ht]
\centering
\caption{Scale-Free Classification of CiteSeer dataset.}
\resizebox{\textwidth}{!}{
\begin{tabular}{|ccc|ccc|ccc|ccc|c|c|}
\hline
\multicolumn{3}{|c}{$\alpha$} & \multicolumn{3}{|c|}{$n_\text{tail}$} & \multicolumn{3}{c|}{$D$} & \multicolumn{3}{c|}{Class} & Subgraph & Model \\
$1 \times$ & $1.3\times$ & $1.5\times$ & $1 \times$ & $1.3\times$ & $1.5\times$ & $1 \times$ & $1.3\times$ & $1.5\times$ & $1 \times$ & $1.3\times$ & $1.5\times$ &  &  \\
\hline
2.21 & 2.44 & 2.50 & 12 & 8 & 15 & 0.01 & 0.12 & 0.07 & \textbf{Weak} & \textbf{Weak} & \textbf{Weak} & 0 & GenCAT \\
2.71 & 2.63 & 2.34 & 73 & 52 & 81 & 0.04 & 0.04 & 0.05 & \textbf{Strong} & \textbf{Strong} & \textbf{Strong} & 1 & GenCAT \\
2.46 & 2.52 & 2.86 & 365 & 68 & 22 & 0.04 & 0.03 & 0.05 & \textbf{Strong} & \textbf{Strong} & Weak & 2 & GenCAT \\
2.34 & 2.39 & 2.62 & 104 & 89 & 107 & 0.02 & 0.03 & 0.03 & Strong & Strong & Strong & 3 & GenCAT \\
2.49 & 2.78 & 2.36 & 145 & 41 & 155 & 0.03 & 0.06 & 0.04 & \textbf{Strong} & Weak & \textbf{Strong} & 4 & GenCAT \\
2.21 & 2.36 & 2.48 & 76 & 56 & 209 & 0.05 & 0.03 & 0.05 & Strong & Strong & Strong & 5 & GenCAT \\ \hline
3.54 & - & - & 26 & - & - & 0.05 & - & - & \textbf{Weak} & - & - & 0 & chung-lu \\
2.90 & - & - & 124 & - & - & 0.05 & - & - & \textbf{Strong} & - & - & 1 & chung-lu \\
3.53 & - & - & 44 & - & - & 0.03 & - & -& Weak & - & - & 2 & chung-lu \\
2.90 & - & - & 189 & - & - & 0.05 & - & - & Strong & - & - & 3 & chung-lu \\
2.99 & - & -& 124 & - & - & 0.03 & - & - & \textbf{Strong} & - & - & 4 & chung-lu \\
7.55 & - & - & 7 & - & - & 0.05 & - & - & \textbf{Weak} & - & - & 5 & chung-lu \\ \hline
3.97 & 3.22 & 3.37 & 24 & 28 & 21 & 0.03 & 0.04 & 0.02 & \textbf{Weak} & \textbf{Weak} & \textbf{Weak} & 0 & SynCo \\
3.83 & 2.68 & 2.32 & 54 & 171 & 193 & 0.06 & 0.04 & 0.06 & Weak & \textbf{Strong} & \textbf{Strong} & 1 & SynCo \\
3.44 & 2.56 & 2.65 & 57 & 294 & 143 & 0.04 & 0.04 & 0.04 & Weak & \textbf{Strong} & \textbf{Strong} & 2 & SynCo \\
4.09 & 2.75 & 3.36 & 50 & 254 & 114 & 0.06 & 0.04 & 0.04 & \textbf{Weak} & Strong & \textbf{Weak} & 3 & SynCo \\
3.93 & 3.19 & 3.23 & 61 & 94 & 83 & 0.05 & 0.03 & 0.02 & Weak & Weak & Weak & 4 & SynCo \\
4.01 & 2.81 & 3.94 & 44 & 163 & 48 & 0.05 & 0.04 & 0.04 & \textbf{Weak} & Strong & \textbf{Weak} & 5 & SynCo \\ \hline
2.70 & - & - & 39 & - & - & 0.05 & - & - & Weak & - & - & 0 & CiteSeer \\
2.55 & - & - & 193 & - & - & 0.05 & - & - & Strong & - & - & 1 & CiteSeer \\
3.00 & - & - & 84 & - & - & 0.04 & - & - & Strong & - & - & 2 & CiteSeer \\
5.98 & - & - & 23 & - & - & 0.05 & - & -& Weak & - & - & 3 & CiteSeer \\
2.89 & - & - & 187 & - & - & 0.03 & - & - & Strong & - & - & 4 & CiteSeer \\
3.72 & - & - & 54 & - & - & 0.04 & - & - & Weak & - & - & 5 & CiteSeer \\
\hline
\end{tabular}
}
\label{tab:sfciteseer}
\end{table*}

For the PubMed dataset, the proposed model achieved significant results, with a loss in classification consistency occurring only when the number of nodes was increased by 1.5×. In comparison to GenCAT, the node degree distribution remained more stable under augmentation. The Chung–Lu algorithm also produced relevant results, however, this model is not inherently designed to increase the number of nodes in the cloned graph.

For the CiteSeer dataset, the best results were obtained when the number of nodes was increased by 1.5×. In the smaller augmentation setting, SynCo was not able to fully capture the scale-free characteristics of certain subgraphs. Nevertheless, when considering all three experimental scenarios jointly, the proposed approach outperformed the baseline algorithms overall.

\section{Discussion and Conclusion}

This paper introduced SynCo, a synthetic community-aware attributed graph generator designed for benchmarking community detection and graph learning algorithms. Unlike existing generators that impose restrictive structural assumptions, SynCo enables explicit control over community sizes, sub-community structures, node degree distributions, inter-community noise, and attribute generation mechanisms. This design allows the creation of graphs with adjustable structural and semantic complexity, supporting systematic evaluation under controlled conditions.

The experimental analysis demonstrates that SynCo enables interpretable manipulation of structural heterogeneity and attribute dispersion. By independently controlling $\rho$, $\alpha_\text{feat}$, and $\alpha_\text{topo}$, the framework allows researchers to progressively increase task difficulty and study how clustering performance degrades under higher structural mixing and feature noise.

To illustrate this capability, DGCluster~\cite{Bhowmick2024dgcluster} was selected as a representative clustering method for parametric evaluation. However, the proposed framework is not restricted to this algorithm and can be used to benchmark any GNN-based or graph learning approach. The controlled experiments reveal that algorithm sensitivity to hyperparameters is strongly influenced by structural conditions, particularly under higher heterophily levels. This highlights the importance of synthetic generators that allow systematic stress-testing beyond fixed real-world datasets.

In the mimicking and augmentation scenarios, SynCo preserves structural properties more consistently than baseline generators. Nevertheless, some limitations remain. The mimicking stage depends on an external community detection algorithm to recover sub-community structures, which introduces sensitivity to the chosen partitioning method. Additionally, the attribute transfer mechanism relies on rank-based matching between original and synthetic nodes. While this strategy preserves structural roles, it partially reduces feature-level generalization, since attributes are derived from the original data rather than being independently generated.

Overall, SynCo provides a flexible and interpretable framework that bridges fully synthetic graph generation and structure-preserving dataset replication. By enabling fine-grained structural control while maintaining empirical plausibility, the model offers a practical tool for advancing research in attributed graph clustering and graph neural networks.

% ------------------------------------------------
% AUTHOR CONTRIBUTIONS
% ------------------------------------------------

\printcredits

% ------------------------------------------------
% REFERENCES
% ------------------------------------------------

\bibliographystyle{cas-model2-names}
\bibliography{references}

\end{document}